\documentclass[conference]{IEEEtran}
\IEEEoverridecommandlockouts
\usepackage{cite}
\usepackage{amsmath,amssymb,amsfonts}
\usepackage{algorithmic}
\usepackage{graphicx}
\usepackage{textcomp}
\usepackage{xcolor}
\usepackage{url}
\usepackage{tabularx}
\usepackage{subfig}
\usepackage{comment}
\usepackage{multirow}

\def\tablename{Table}
\usepackage{etoolbox}
\makeatletter
\patchcmd{\@makecaption}
  {\scshape}
  {}
  {}
  {}
\makeatletter
\patchcmd{\@makecaption}
  {\\}
  {.\ }
  {}
  {}
\makeatother
\def\tablename{Table}
\def\BibTeX{{\rm B\kern-.05em{\sc i\kern-.025em b}\kern-.08em
    T\kern-.1667em\lower.7ex\hbox{E}\kern-.125emX}}
\begin{document}


\title{Exploring operation parallelism vs. ion movement in ion-trapped QCCD architectures\\
}

\author{\IEEEauthorblockN{Anabel Ovide and Carmen G. Almudever}

\IEEEauthorblockA{\textit{Computer Engineering Department, Universitat Politècnica de València, Valencia, Spain}}}

\maketitle

\begin{abstract}

Ion-trapped Quantum Charge-Coupled Device (QCCD) architectures have emerged as a promising alternative to scale single-trap devices by interconnecting multiple traps through ion shuttling, enabling the execution of parallel operations across different traps. While this parallelism enhances computational throughput, it introduces additional operations, raising the following question: do the benefits of parallelism outweigh the potential loss of fidelity due to increased ion movements?

This paper answers this question by exploring the trade-off between the parallelism of operations and fidelity loss due to movement overhead, comparing sequential execution in single-trap devices with parallel execution in QCCD architectures. We first analyze the fidelity impact of both methods, establishing the optimal number of ion movements for the worst-case scenario. Next, we evaluate several quantum algorithms on QCCD architectures by exploiting parallelism through ion distribution across multiple traps. This analysis identifies the algorithms that benefit the most from parallel executions, explores the underlying reasons, and determines the optimal balance between movement overhead and fidelity loss for each algorithm.

\end{abstract}


\section{Introduction}

The present era is characterized by an ongoing competitive landscape between multiple quantum technologies pursuing the objective of achieving fault-tolerant quantum computing~\cite{Kjaergaard_2020,Huang2020,doi:10.1080/09500340008244052,Henriet_2020,HAFFNER2008155,Bruzewicz_2019}. Each of these technologies offers distinctive advantages, yet they all face similar limitations, including scalability—a challenge addressed by some through the exploration of modularity~\cite{10.1145/3457388.3458674, 9923784}. This paper focuses on ion trap processors, which have embraced a modular approach known as the Quantum Charge-Coupled Device (QCCD)~\cite{Pino_2021,PhysRevX.13.041052,quantinuumScalability}
, with future expansions already in development~\cite{quantinuumRoadmap}.

One highlight of QCCD architectures, in addition to their scalability, is their ability to execute parallel operations across different trapping zones. Although parallel execution is feasible in a single trap, it leads to crosstalk, reducing fidelity~\cite{Figgatt2019}, an issue overcome by the QCCD parallelization approach. Therefore, parallelism could be enhanced in the compilation process by distributing ions that can be operated concurrently in separate traps. However, prior work on compilations has primarily focused on minimizing ion movement to preserve fidelity\cite{schoenberger2024shuttling,Saki_2022,schoenberger2023using,Schmale_2022}, and the impact of prioritizing parallel execution over movement reduction remains unexplored.

While enhancing parallelism can lead to increased computational throughput by allowing concurrent execution of operations and reducing overall execution time, it incorporates additional SWAP and shuttling operations for ion movement, raising the following question: do the benefits of parallelism outweigh the potential loss of fidelity due to the increased move of ions?

This paper fills this gap by analyzing the trade-off between parallelism of operations and loss of fidelity due to movement overhead. It is structured as follows: Section II provides an overview of the QCCD technology, highlighting the compilation process while considering operation parallelism. In Section III, fully parallelizable random quantum circuits are executed on both a single-trap device (sequential execution) and a QCCD architecture (parallel execution) to compare their performance under worst-case conditions. In the last Section IV, a set of well-known quantum algorithms is evaluated in QCCD and single-trap architectures, determining the optimal number of ion movements when executed in parallel by varying the number of traps and ions per trap. The paper ends with conclusions and an outline of future work.

\section{Compilation in QCCD architectures}
The Quantum Charge-Coupled Device (QCCD) architecture presents the advantage of scalability and parallel execution of operations across different traps without introducing crosstalk.  QCCD consists of multiple traps connected by a shuttle mechanism~\cite{Harty_2014}, which enables the physical transfer of ions between them. To perform two-qubit gates, involved ions must be located within the same trap, requiring ion movements—such as shuttling and SWAP operations. In particular, ions need to be positioned at the ends of the ion chain using SWAP operations before being shuttled to a different trap. However, these operations are costly and can degrade execution fidelity.


It is essential to minimize these movement operations in the compilation process. Current QCCD compilation techniques focus on reducing such operations; however, none of them exploits the potential for parallelism. To do so, some steps of the compilation process should be modified. The initial placement of qubits~\cite{10821469,upadhyay2022shuttleefficient}—which assigns virtual qubits (qubits in a quantum circuit) to physical qubits (ions)—is typically designed to place interacting qubits in close proximity (i.e., within the same trap or in adjacent traps). However, when considering parallelism, one should focus not only on the proximity of interacting qubits but also on distributing qubits across different traps to enhance parallelism. 

\begin{figure}[t] 
\vspace{-8pt}
    \subfloat[]{
        \includegraphics[width=0.49\columnwidth]{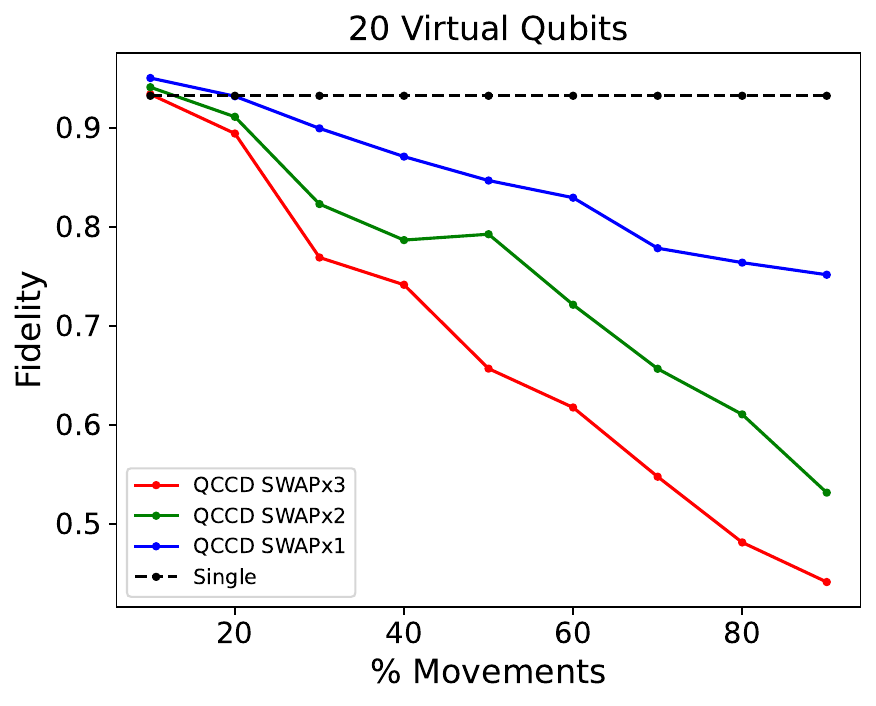}
        }
    \subfloat[]{
           \includegraphics[width=0.49\columnwidth]{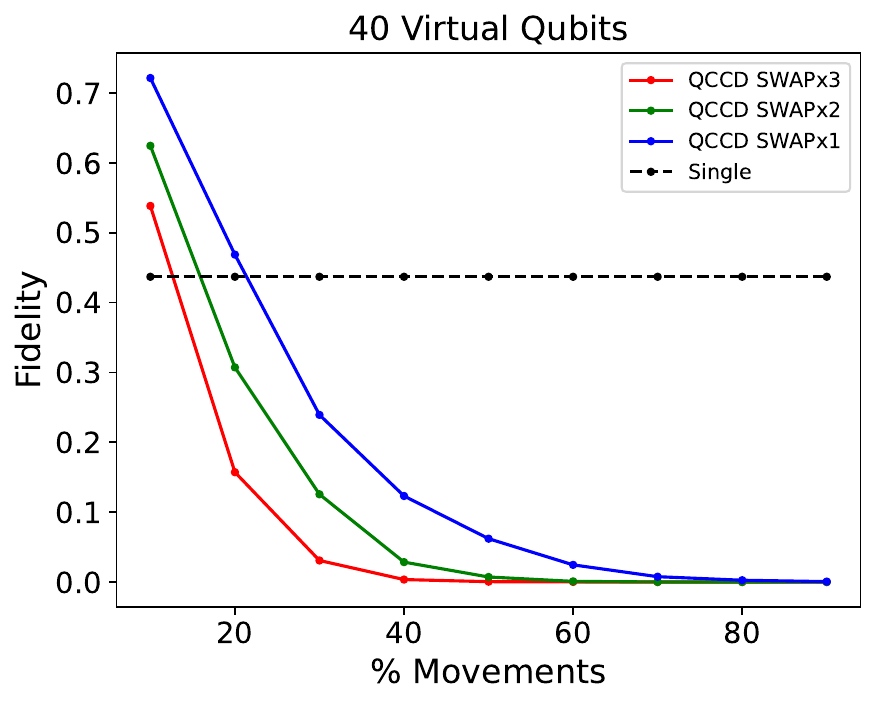}
    }
    \vspace{-5pt}
    \caption{\small{Circuit fidelity vs. percentage of ions movement. A 100\% indicates maximum ion movements during execution (all qubits are moved whenever they participate in a new operation). Different SWAP fidelities are considered: 3,2, and 1 times worse than a CNOT gate. A fully sequential execution is represented with a black dashed line.}}
    \label{fig:back}
    \vspace{-11pt}
\end{figure}

The next compilation step involves qubit routing and scheduling~\cite{https://doi.org/10.4230/lipics.tqc.2019.5}, which moves ions between traps to execute two-qubit gates. Current strategies mainly aim to minimize the number of SWAPs and ion shuttling while anticipating future operations~\cite{schoenberger2024shuttling,Saki_2022,schoenberger2023using,Schmale_2022}. However, when parallelism is considered, a new trade-off between minimizing operations and effectively distributing ions among different traps emerges.

\begin{figure}[h] 
     \centering
    \subfloat[]{
        \includegraphics[width=0.60\columnwidth]{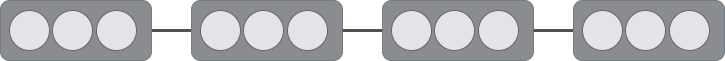}
        }
    \vspace{-5pt}
    \caption{{1D-linear QCCD with 4 traps and 3 ions per trap}}
    \label{fig:Topology}
    \vspace{-8pt}
\end{figure}

By comparing existing compilation techniques with an improved approach that incorporates parallelism, one can observe a consistent tension between reducing movements and maximizing parallelism. This suggests the need to determine the optimal balance between parallelism and ion movements in the compilation strategy. Fig.~\ref{fig:back} compares the fidelity of fully parallelizable random circuits when executed using two different approaches: a parallel implementation within a 1D-linear QCCD architecture (Fig.~\ref{fig:Topology}) and a sequential execution in a single-trap device. The results reveal two key points: (i) a small number of qubits yields better performance in a sequential execution within a single-trap device (Fig. 1a); however, as the number of qubits increases (Fig. 1b), a parallel execution can improve fidelity if the number of ion movements remains manageable; and (ii) as expected, improving SWAPs fidelity results in an increase of circuit fidelity when considering parallel execution. These results underscore the importance of parallelism as the number of qubits increases.


 \section{Analysis of the worst-case scenario}

This section explores the trade-off between operation parallelism and ion movement overhead by analyzing the worst-case scenario. Specifically, it compares the execution of a set of fully parallelizable random circuits on a single-trap device (sequential execution) and on a QCCD architecture (parallel execution) when the ion movements are maximal.

\subsection{Framework}

A comparison is conducted between a sequential execution in a single-trap device and a parallel execution in a QCCD architecture. To this end, a random, fully parallelizable benchmark circuit set has been developed for a 1D-linear array QCCD topology with a trap capacity of 3 ions (Fig.~\ref{fig:Topology}). At the beginning of the execution, each trap contains only two ions, leaving one free space to facilitate ion movements between traps. 

The benchmarks are random quantum circuits composed of an even number of virtual qubits with a circuit depth corresponding to the number of qubits (e.g., a circuit with 20 virtual qubits has a depth of 20). Additionally, the benchmarks allow varying degrees of ion movements through two-qubit gates distribution across different qubit pairs—a movement percentage of 0\% indicates that no SWAP or shuttling operations are needed, whereas 100\% means all qubits are moved to a different trap in each time step (discrete unit of time in where quantum gates are executed concurrently on different qubits). Full parallelization is achieved by ensuring that at each time step, every qubit pair performs a two-qubit gate operation.

To achieve a fully parallel execution, a compilation process aware of operation parallelization is essential. 
A naive qubit routing and scheduling algorithm have been developed to address this need while using the Spatio-Temporal Aware (STA) algorithm~\cite{10821469} for the initial qubit assignment. After allocating each pair of interacting qubits in a different trap (i.e. two qubits can be placed per trap), all two-qubit gates are simultaneously executed at each time step. Between time steps, the naive algorithm performs the required ion shuttling and SWAP operations to relocate interacting qubits in their corresponding traps. It should be noted that this qubit routing and scheduling algorithm does not consider future operations; thereby, the possible optimization in reducing movement operations is minimal. 

This approach allows the study of the worst-case scenario by using unstructured circuits, and maximizing operation parallelization but also ion movements. To conduct the experiments, the simulator framework QCCDSim~\cite{murali2020architecting} was employed.  Two performance metrics were evaluated: circuit fidelity and qubit coherence after execution, which is calculated as follows:
\begin{equation}
    C = e^{-\frac{t}{T_2}}
    \label{eqn:coherence}
\end{equation}

where $t$ denotes the execution time and $T_2$ represents the qubit coherence time. QCCDSim provides a fidelity metric that contemplates error gate, ion chain vibrational motion, and fidelity reduction due to shuttling. This metric has been further refined by incorporating the effects of decoherence, achieved by multiplying it by equation~\ref{eqn:coherence}. For further details on the computation of fidelity, refer to~\cite{murali2020architecting}.

\subsection{Results}

\begin{figure}[t] 
\vspace{-8pt}
    \subfloat[]{
        \includegraphics[width=0.49\columnwidth]{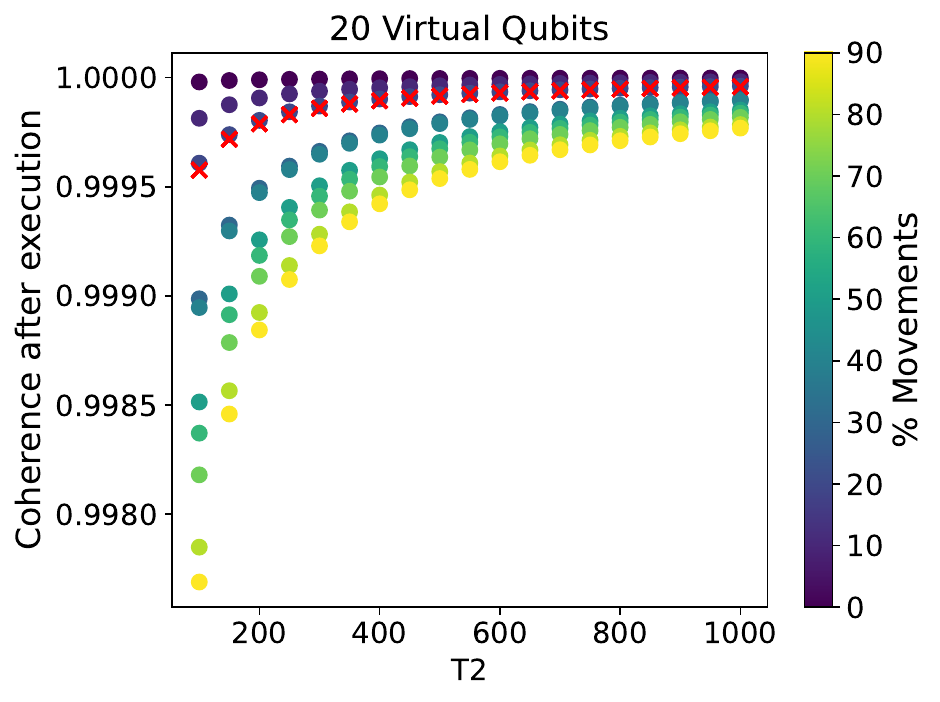}
        }
    \subfloat[]{
           \includegraphics[width=0.49\columnwidth]{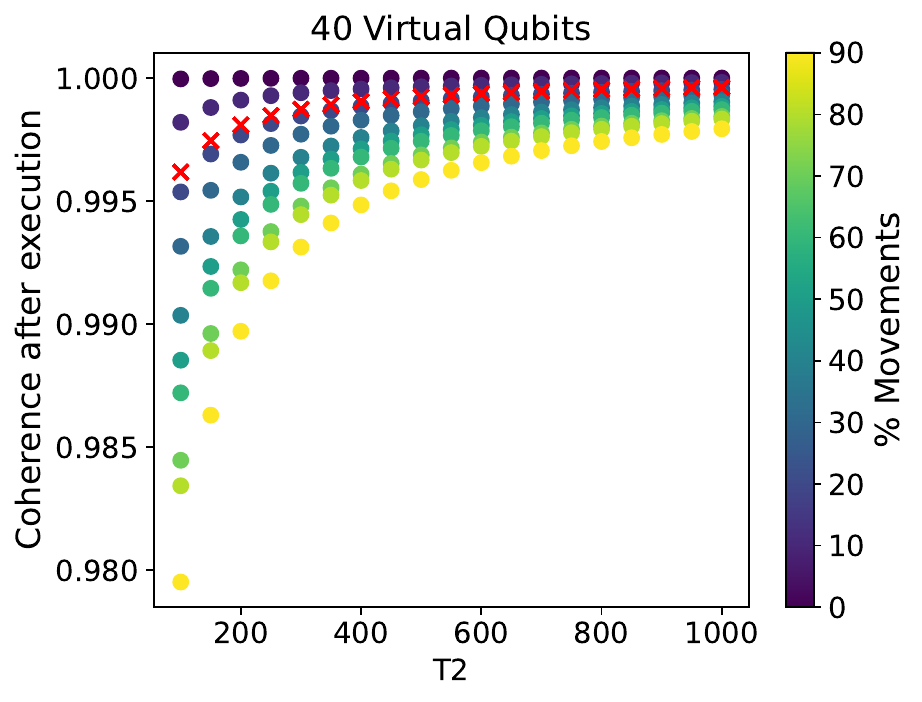}
    } \\
        \subfloat[]{
        \includegraphics[width=0.5\columnwidth]{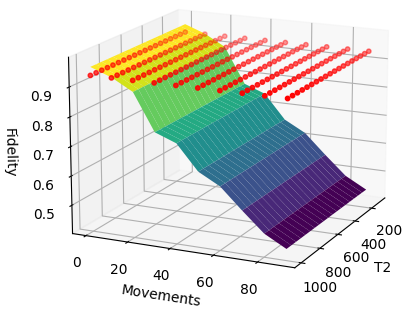}
        } 
    \subfloat[]{
           \includegraphics[width=0.5\columnwidth]{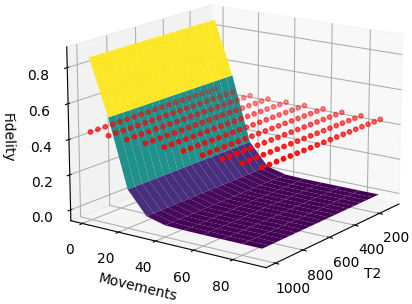}
    }
    \vspace{-5pt}
    \caption{\small{ (a), (b) Coherence and (c), (d) circuit fidelity vs. \% of ion movements and T2 (ms), after execution in a 1D-linear QCCD topology. 20 and 40 virtual qubits are considered. Sequential execution in a single-trap device is illustrated as red dots or crosses.}}
    \label{fig:3D}
    \vspace{-11pt}
\end{figure}

The experiments were conducted using the previously described benchmarks, and the naive qubit routing and schedule method on both: a single-trap device and a 1D-linear QCCD architecture (Fig.~\ref{fig:Topology}). Fig.~\ref{fig:3D} illustrates the results for 20 and 40 virtual qubits in which traps are adjusted in accordance with the virtual ones (e.g. 20 virtual qubits, 10 traps with a capacity of 3 ions each). Coherence times ($T_2$) ranging from 200ms to 1000ms were considered~\cite{IonQ,Myerson_2008}. The percentage of ion movements varied from 0\% to 100\%.

Results indicate that the primary factor impacting fidelity is the number of movement operations: more movements lead to lower fidelity due to increased SWAP and ion shuttling operations. Qubit decoherence had a relatively smaller impact on performance compared to the number of operations.  In circuits comprising a substantial number of virtual qubits (i.e. 40), parallel execution outperforms sequential execution, provided that movement overhead is approximately 20\% or below.

These findings suggest that algorithms requiring a large number of movement operations may benefit the most from a sequential execution approach or from the use of QCCD devices with a reduced number of traps and larger trap capacities. In contrast, algorithms with lower ion movements rates achieve better fidelity with parallel execution.

However, structured quantum algorithms are expected to show different results, as fewer movement operations would be required, and compilation techniques would play a more significant role in reducing these movements as we will explore in the next section. 

\section{Impact of Parallelism on Quantum Algorithm Performance}

After analyzing the worst-case scenario, a set of well-known quantum algorithms was executed in the QCCD architecture. The results were compared against sequential execution, identifying optimal QCCD topologies that will enhance parallelism for different structured algorithms.

\subsection{Experimental setup}

A selection of structured quantum algorithms—including the Cuccaro Adder (CA), Draper Adder (DA), Quantum Approximate Optimization Algorithm (QAOA), and Quantum Fourier Transform (QFT)—was made based on their different structures and their use in previous works~\cite{10821469,Saki_2022,upadhyay2022shuttleefficient,murali2020architecting}. Table~\ref{tab:Benchmarks} summarizes the main characteristics of each algorithm, where TS refers to the time step, and in which the average of the movement of ions (shuttling) per time step in the worst-case scenario is calculated as follows.

\begin{equation}
    \textit{Av.}_{\text{ion\_mov/TS}} = \sum_{i}^N\frac{M_{q_i}}{D}
\end{equation}

where $N$ is the total number of qubits, $q_i$ represents a virtual qubit, $D$ the depth of the algorithm, and $M_{q_i}$ is the number of movements per qubit (one per interaction with a different qubit).

CA has the fewest two-qubit gates but the highest circuit depth, having few ion movements compared with other algorithms, making it the most sequential of the selected algorithms (also represented in the average gates per time-step column). In contrast, QAOA and QFT feature a greater number of two-qubit gates, movement operations, and shallower depths, making them more parallelizable algorithms. DA falls between CA and QAOA/QFT.

\begin{table}[t!]
    \centering
    \vspace{-3pt}
    \caption{Benchmarks-40 Virtual Qubits}
    \label{tab:Benchmarks}
    \vspace{-8pt}
    \begin{tabular}{*{5}{c}}
        \multirow{2}{*}  & \multirow{2}{*}{\textbf{Depth}} & \multirow{2}{*}{\textbf{2q Gates}} & \multirow{2}{*}{\textbf{Av. 2q-Gates/TS}} &\multirow{2}{*}{\textbf{Av. Shuttle/TS}}\\
        & & & \\ [-1ex]
        \hline
        \textbf{CA}  & 283 & 321 & 1.13  &1.28\\
        \textbf{DA}  & 113 & 590 & 5.22 &9.91\\
        \textbf{QAOA}  & 77 & 780 & 10.13 & 19.74\\
        \textbf{QFT}  & 77 & 780 & 10.13 & 19.74\\
        \hline
    \end{tabular}
    \vspace{-8pt}
\end{table}

The selected algorithms were executed in the previously mentioned QCCDSim simulator~\cite{murali2020architecting}, compiled using the Spatio-Temporal Aware (STA) initial placement algorithm~\cite{10821469} and the qubit routing and scheduling algorithm proposed by Saki et al.~\cite{Saki_2022}. These compilation algorithms do not consider parallelism; instead, they are designed to minimize ion movements. Consequently, the parallelism observed in the experiments 
was achieved through the configuration of different QCCD topologies. Specifically, the degree of parallelism was controlled by adjusting the number of traps and the number of ions per trap. An increase in the number of traps with a reduction in the number of qubits per trap results in a higher parallelism. Conversely, reducing the number of traps with an increase in the number of qubits per trap leads to more sequential execution.

\subsection{Experiments}

\begin{table*}[ht]
    \centering
    \vspace{-3pt}
    \caption{Optimal QCCD Architectures comparison}
    \label{tab:Cmparison}
     \vspace{-5pt}
    \begin{tabular}{*{11}{c}}
        \cline{2-10}
\multirow{-1}{*}{} & \multicolumn{3}{|c|}{\textbf{20 Virtual Qubits}} & \multicolumn{3}{c|}{\textbf{40 Virtual Qubits}}& \multicolumn{3}{c|}{\textbf{50 Virtual Qubits}} \\
        \cline{2-10}
        
        \multirow{2}{*} & \multirow{2}{*}{\textbf{Opt. Traps}} & \multirow{2}{*}{\textbf{Max. Traps}} & \multirow{2}{*}{\textbf{$\Delta F(\%)$}} & \multirow{2}{*}{\textbf{Opt. Traps}} & \multirow{2}{*}{\textbf{Max. Traps}} & \multirow{2}{*}{\textbf{$\Delta F(\%)$}} & \multirow{2}{*}{\textbf{Opt. Traps}} & \multirow{2}{*}{\textbf{Max. Traps}} & \multirow{2}{*}{\textbf{$\Delta F(\%)$}}\\
        & & & & & & \\ [-1ex]
        \hline
        \textbf{CA} & 2 & 3 & \text{2.24\%}& 4 & 20 & \textbf{29.9\%}& 6 & 25 & \textbf{58.11\%}\\
        \textbf{DA} & 2 & 2 & \text{0.25\%}  & 2 & 5 & \textbf{36.89\%}& 3 & 7 & \textbf{99.5\%}\\
        \textbf{QAOA}  & 2 & 2 & \text{0.95\%} & 2 & 4 & \textbf{48.44\%}& 3 & 4 & \textbf{119.48\%} \\
        \textbf{QFT}  & 2 & 2 & \text{0.77\%} & 2 & 4 & \textbf{48.63\%}& 3 & 4 & \textbf{159.78\%}  \\
        \hline
    \end{tabular}
    \vspace{-7pt}
\end{table*}

The experiments were conducted using a 1D-linear array topology (Fig.~\ref{fig:Topology}). Trap capacities were adjusted to accommodate the virtual qubits, plus one extra space to facilitate ion movements. The number of traps varied while the total number of physical qubits was kept constant, thus modifying qubit distribution. 

\begin{figure}[t] 
\vspace{-7pt}
    \subfloat[]{
        \includegraphics[width=0.49\columnwidth]{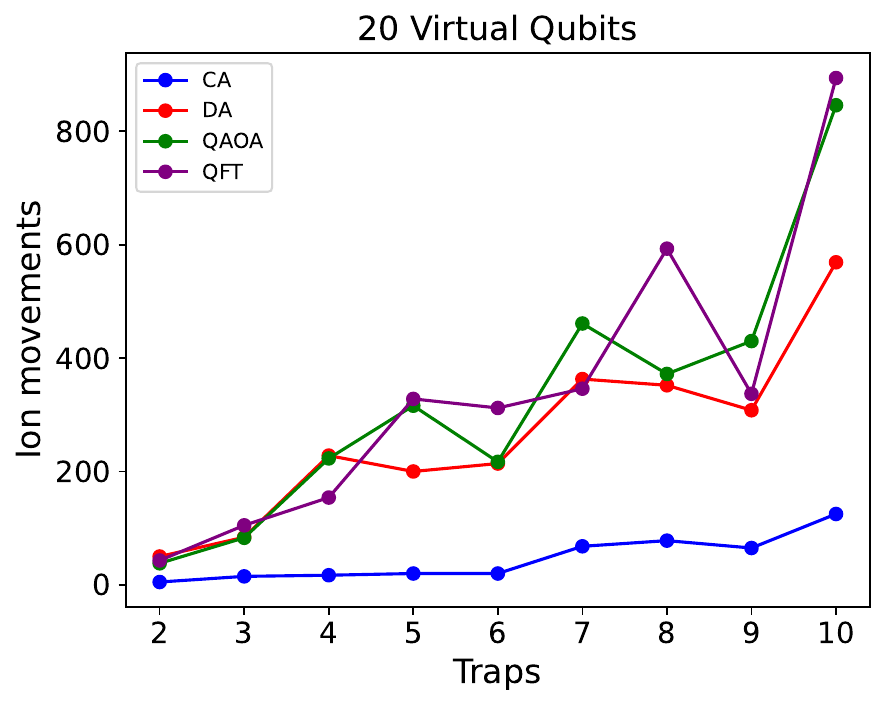}\label{fig:sub1}
        }
    \subfloat[]{
           \includegraphics[width=0.49\columnwidth]{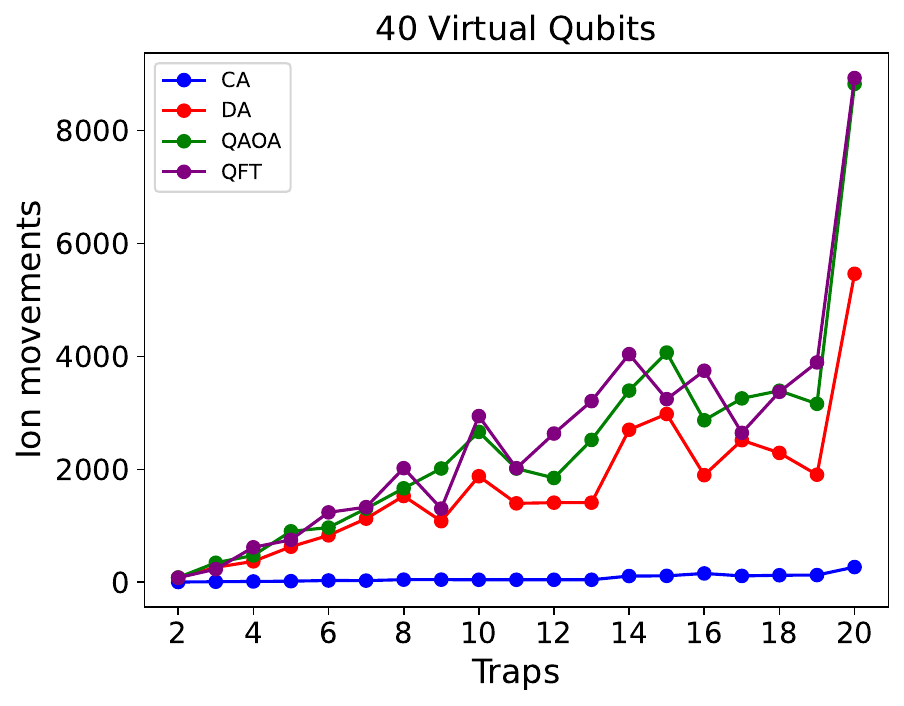}\label{fig:sub2}
    }\vspace{-10pt}\\
    \subfloat[]{
        \includegraphics[width=0.49\columnwidth]{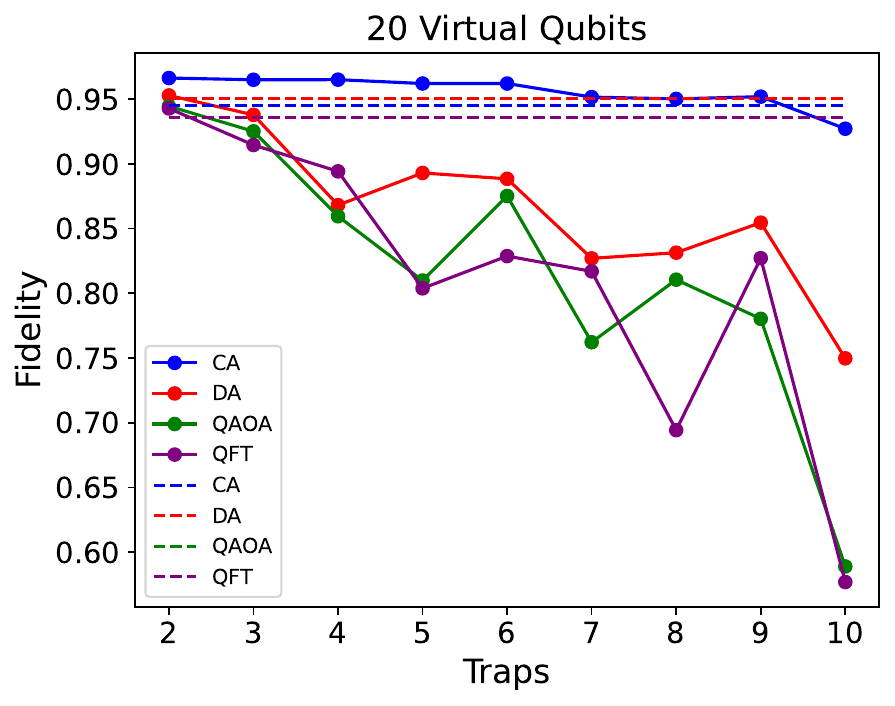}\label{fig:sub3}
        }
    \subfloat[]{
           \includegraphics[width=0.49\columnwidth]{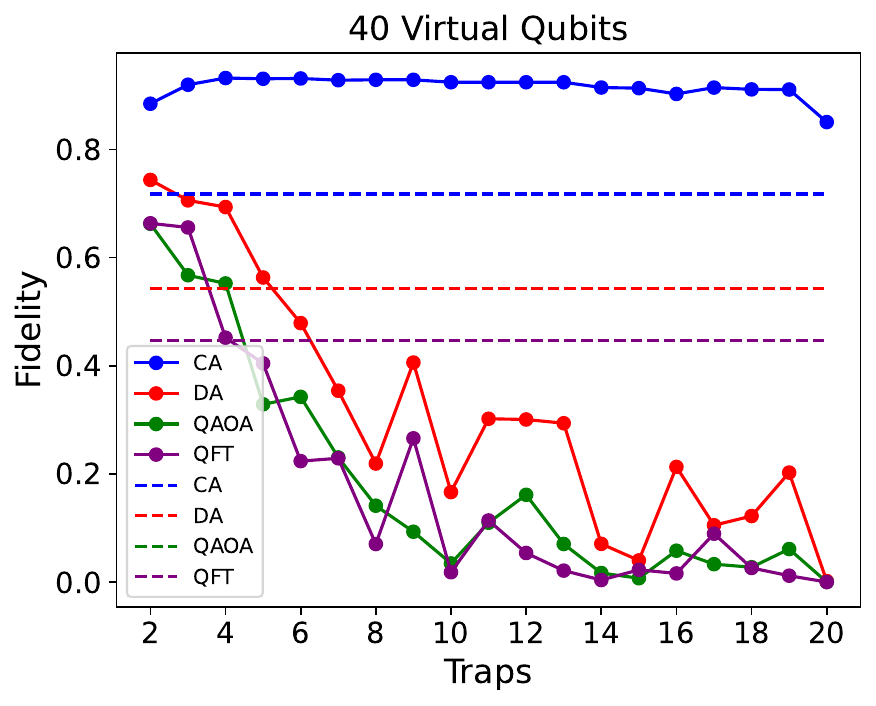}\label{fig:sub4}
    }
    \vspace{-5pt}
    \caption{\small{Comparison between a set of different quantum algorithms executing 20 (a) and 40 (b) virtual qubits in a 1D-linear QCCD topology. Dashed lines represent sequential execution in a single-trap device.}}
    \label{fig:QA}
    \vspace{-10pt}
\end{figure}

Fig.~\ref{fig:QA} shows that parallelism becomes more beneficial as the number of virtual qubits increases, as illustrated in the previous section. The parallel execution of the CA algorithm outperforms the sequential one in approximately half of the cases with 20 virtual qubits and in all cases with 40 virtual qubits. This is due to the limited number of qubit interactions (movement of ions), consistent with the scenario described in the previous section, where, on the basis of the reference of 20\% or fewer ion movements, parallel execution outperforms the sequential one, allowing a modest level of parallelization with few additional operations. Furthermore, reducing the number of ions per trap improves the fidelity of gate operations.

For the remaining quantum algorithms, as the number of virtual qubits increases, certain levels of parallelism surpass sequential execution in terms of fidelity. Despite increasing ion movements, the enhanced parallelization achieved by expanding the number of traps improves computational throughput, thereby enhancing fidelity. Across all algorithms, the best results are typically achieved with a configuration of a few traps, which strikes a balance between enabling some parallelism while not raising the movement overhead considerably. Note that QFT and QAOA dash lines (single-trap) overlap.

Table~\ref{tab:Cmparison} presents the best-performing QCCD architectures for executing quantum circuits with 20, 40, and 50 qubits. It compares three key metrics: the maximum number of traps (Max. Traps) in which a parallel execution outperforms a sequential one; the optimal number of traps (Opt. Traps), representing the ideal topology that maximizes fidelity; and the improvement in fidelity ($\Delta F(\%)$) achieved under the optimal number of traps. The data show that as the number of virtual qubits increases, so does the optimal and maximum number of traps, as well as the fidelity improvement from parallel execution. This highlights that parallelism is crucial for scalability: as the number of virtual qubits in an algorithm increases, the benefits of parallel execution become more pronounced.

From these results, three key points can be derived: (i) QCCD technology is essential for scalability, as it demonstrates that increasing qubits favor the adoption of parallel execution. (ii) Despite the fact that the compilation approaches utilized in these experiments are not specifically designed for parallelization, certain instances have demonstrated superior fidelity compared to their sequential counterparts. This suggests that employing compilation techniques that consider parallelism could yield even better results. (iii) Quantum algorithm parameters can be used to identify the balance between ion movements and parallelism that maximizes their performance (fidelity). This, in turn, also helps to determine the optimal topology that maximizes fidelity.

\section{Conclusions and future work}



This paper has explored the trade-off between ions movement when maximizing parallelism and the fidelity loss, leading to the following conclusions: (i) Exploiting the parallelism of quantum algorithms is more beneficial when the resulting movement overhead is below a certain threshold. (ii) When scaling the algorithms to a higher number of qubits, a parallel approach tends to outperform a sequential execution. (iii) Compilation techniques for QCCD architectures can be improved by considering the parallelism-communication overhead trade-off. (iv) Optimal topologies (number of traps and ions per trap) can be derived for different quantum algorithms on the basis of their characteristics. 

In future work, we will develop compilation methods to fully leverage the benefits of parallelism.

\section*{Acknowledgments}

The authors acknowledge financial support from the European Union’s Horizon Europe research and innovation program through the project Quantum Internet Alliance under grant agreement No. 101102140. CGA also acknowledges support from the Spanish Ministry of Science, Innovation and Universities through the Beatriz Galindo program 2020 (BG20-00023) and the European ERDF under grant PID2021-123627OB-C51.

\bibliographystyle{ieeetr}
\bibliography{bib/main}

\end{document}